# Giant nonlinear optical responses from photon avalanching nanoparticles

Changhwan Lee[1], Emma Xu[1], Yawei Liu[2,3], Ayelet Teitelboim[2], Kaiyuan Yao[1], Angel Fernandez-Bravo[2], Agata Kotulska[4], Sang Hwan Nam[5], Yung Doug Suh[5,6,*], Artur Bednarkiewicz[4,*], Bruce E. Cohen[2,7,*], Emory M. Chan[2,*], and P. James Schuck[1,*]

[1]Department of Mechanical Engineering, Columbia University, New York, NY, United States.
[2]The Molecular Foundry, Lawrence Berkeley National Laboratory, Berkeley, California 94720, United States.
[3]State Key Laboratory of Rare Earth Resource Utilization, Changchun Institute of Applied Chemistry, Chinese Academy of Sciences, 130022, Changchun, China.
[4]Institute of Low Temperature and Structure Research, Polish Academy of Sciences, 50-422 Wroclaw, Poland.
[5]Laboratory for Advanced Molecular Probing (LAMP), Korea Research Institute of Chemical Technology (KRICT), DaeJeon 305-600, South Korea.
[6]School of Chemical Engineering, Sungkyunkwan University (SKKU), Suwon 440-746, South Korea.
[7]Division of Molecular Biophysics & Integrated Bioimaging, Lawrence Berkeley National Laboratory, Berkeley, California 94720, United States.
*p.j.schuck@columbia.edu; emchan@lbl.gov; becohen@lbl.gov; a.bednarkiewicz@intibs.pl; ydsuh@krict.re.kr

## Abstract

**Avalanche phenomena leverage steeply nonlinear dynamics to generate disproportionately high responses from small perturbations and are found in a multitude of events and materials[1], enabling technologies including optical phase-conjugate imaging,[2] infrared quantum counting,[3] and efficient upconverted lasing[4-6]. However, the photon avalanching (PA) mechanism underlying these optical innovations has been observed only in bulk materials and aggregates[6,7], and typically at cryogenic temperatures[5-8], limiting its utility and impact in many applications. Here, we report the realization of PA at room temperature in single nanostructures – small, $Tm^{3+}$-doped upconverting nanocrystals – and demonstrate their use in superresolution imaging at wavelengths that fall within near-infrared (NIR) spectral windows of maximal biological transparency. Avalanching nanoparticles (ANPs) can be pumped by either continuous-wave or pulsed lasers and exhibit all of the defining features of PA. These hallmarks include clear excitation power thresholds, exceptionally long rise time at threshold, and a dominant excited-state absorption that is >13,000 times larger than ground-state absorption. Beyond the avalanching threshold, ANP emission scales nonlinearly with the 26th power of pump intensity, resulting from induced positive optical feedback in each nanocrystal. This enables the experimental realization of photon-avalanche single-beam superresolution imaging (PASSI)[7], achieving sub-70 nm spatial resolution using only simple scanning confocal microscopy and before any computational analysis. Pairing their steep nonlinearity with existing superresolution techniques and computational methods[9-11], ANPs allow for imaging with higher resolution and at ca. 100-fold lower excitation intensities than is possible with other probes. The low PA threshold and exceptional photostability of ANPs also suggest their utility in a diverse array of applications[7] including sub-wavelength bioimaging[7,12,13], IR detection, temperature[14-16] and pressure[17] transduction, neuromorphic computing[18], and quantum optics[19,20].**

## Main

The primary advantage of PA is its combination of extreme nonlinearity and efficiency, achieved without any periodic structuring or interference effects. PA was first observed in $Pr^{3+}$-doped bulk crystals, which exhibited a sudden increase in upconverted luminescence when excited beyond a critical pump laser intensity ($I_P$)[3]. Its discovery quickly led to the development of other lanthanide-based bulk PA materials, utilized e.g. in efficient upconverted lasers[4-6,21]. PA has been observed mostly at cryogenic temperatures[5-8], with a few room-temperature exceptions[6,7,22-25], but its unique properties continue to spark interest over diverse fields[6,7].

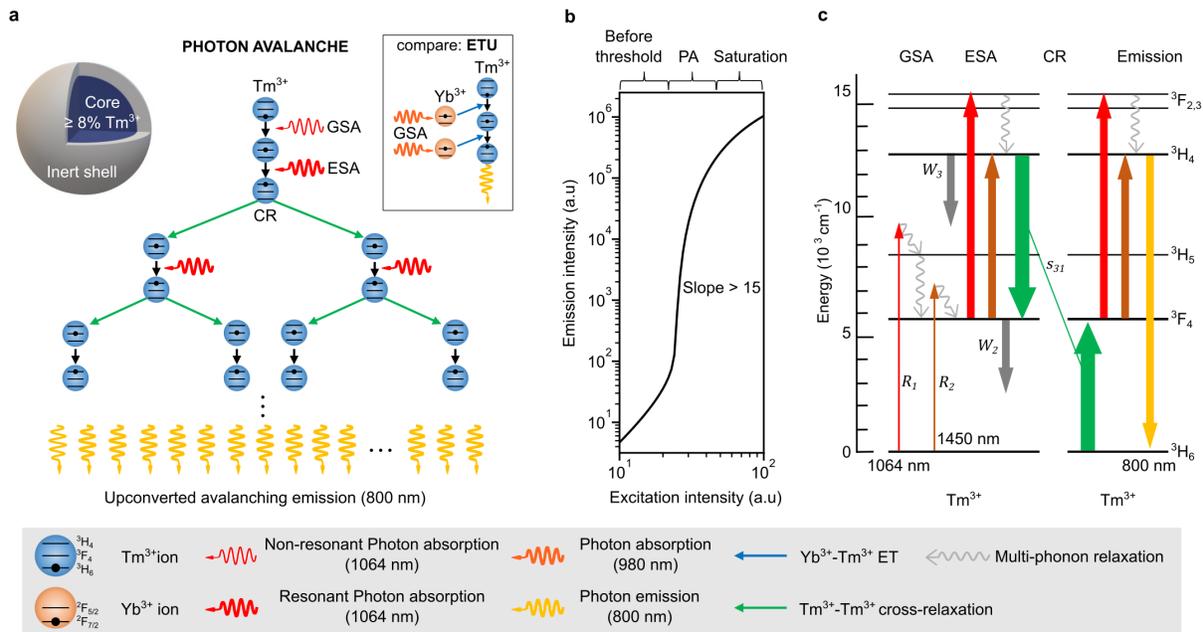

**Fig. 1. Photon avalanching mechanism in $Tm^{3+}$-doped nanocrystals. a,** Schematic of core/shell ANPs, with avalanching occurring when core $Tm^{3+} \geq 8\%$. (*Inset*) Standard ETU process, in which $Yb^{3+}$ ions sensitize ground state absorption, precluding PA. **b,** Model plot of emission intensity vs. excitation intensity, showing the three stages of PA behavior. **c,** Energy levels of the $4f^{12}$ manifolds of $Tm^{3+}$. $R_1$, $R_2$ = ground- and excited-state excitation rates, respectively. $W_2$, $W_3$ = $^3F_4$ and $^3H_4$ decay rates, respectively. CR = cross-relaxation. $s_{31}$ = CR rate.

PA is a positive feedback system[6] analogous to the second order phase transition of ferromagnetic spin systems, comparisons that have proven useful for modeling the process[5,26]. In lanthanide-based PA, a single ground-state absorption (GSA) event initiates a chain reaction of excited-state absorption (ESA) and cross-relaxation events between lanthanide ($Ln^{3+}$) ions, resulting in the emission of many upconverted photons (Fig. 1a). This mechanism amplifies the population of excited states, such as the 800-nm-emitting $Tm^{3+}$ $^3H_4$ level (Fig. 1c), through a positive feedback loop of ESA from an intermediate state ($^3F_4$) followed by cross-relaxation back down to the same intermediate state while exciting a second ground-state ion up to its intermediate state. This process can effectively double the $^3F_4$ population, and repeated looping results in nonlinear amplification of excited state populations.

The sensitivity of $Ln^{3+}$ photophysics to local material properties has precluded the realization of PA in nanomaterials and has hindered room temperature operation. Avalanche-like behavior in previous nanoparticle designs was ultimately the result of the formation of larger aggregate materials[8], non-PA thermal mechanisms[27,28], or of pre-avalanche energy-looping (EL)[6,12,14,29-36], with nonlinear order *s* ranging from 2-7 (*s* is defined by $I_E = I_P^s$, where $I_E$ is emission intensity)[7,12,30]. There remains strong motivation for developing PA in nanoparticles, as they offer the benefits of tunable, solution-processed synthesis, assembly, and incorporation into varied device platforms, novel nanotechnologies and unique environments[31,37] using biocompatible surface chemistries[37-40] and materials[41].

To design nanocrystals that may be capable of PA, we combined four key innovations. The first is the recent design paradigm for $Ln^{3+}$-based upconverting nanoparticles (UCNPs), in which high $Ln^{3+}$ content, engineered energy confinement, and reduced surface losses result in exceptional efficiencies and brightness[31,38,42-47]. A second feature is the choice of $Tm^{3+}$ (Fig. 1a), an ion with a particularly slow intermediate-state decay rate $W_2$, which strongly influences PA behavior[5-7] (see below). The third critical aspect exploits the compositional strategy employed previously for energy looping nanoparticles (ELNPs)[30], in which typical $Yb^{3+}$ sensitizers are omitted and high concentrations of $Tm^{3+}$ ions are doped into a β-phase $NaYF_4$ matrix, enhancing $Tm^{3+}$-$Tm^{3+}$ cross-relaxation and ESA while reducing GSA (Fig. 1). The fourth key element, also shared with ELNPs, is the selection of excitation wavelengths in the NIR-II transparency window (either 1064 nm or 1450 nm; Fig. 1), which are optimized for resonant ESA while maintaining non-resonant GSA, in contrast to the usual wavelengths used for pumping $Tm^{3+}$ (800 nm, or 980 nm when combined with $Yb^{3+}$ sensitization; Fig. 1)[6,12,13,29,48].

To determine if these design criteria enable nanocrystals to host PA, we synthesized $Tm^{3+}$-doped β-$NaYF_4$ core/shell structures 16-33 nm in total diameter[38,42]. In each ANP, the $Tm^{3+}$-doped core is surrounded by an optically inert shell to minimize surface losses[42] (Figs. 1, S1,2, and Tables S1,2). These nanoparticles may be excited in the NIR-II region to emit in the NIR-I region at 800 nm[30]. Both spectral windows are valuable for imaging with limited photodamage through living systems or scattering media[49]. To determine whether PA occurs, we examined them for three definitive criteria[5,6]: (i) stronger pump-laser-induced ESA compared to GSA, with the ratio of ESA to GSA rates exceeding $10^4$ ($R_2/R_1$ in Fig. 1c)[30]; (ii) a clear excitation power threshold, above which a large nonlinear increase in excited state population and emission is observed; and (iii) a slowdown of the excited-state population rise-time at threshold. For PA, rise times typically reach >100× the lifetime of the intermediate state, up to seconds[6]. Together, these criteria delineate

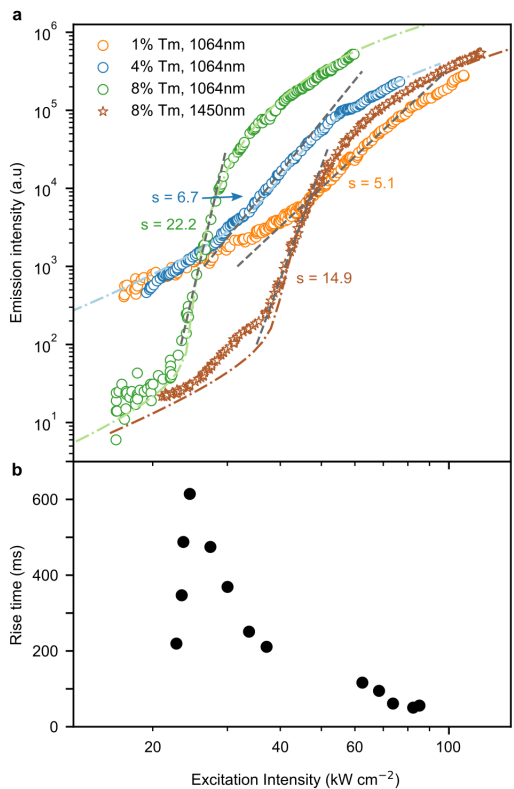

**Fig. 2. Demonstration of nanoparticle photon avalanching. a,** 800 nm emission intensity vs. excitation intensity for 1%, 4%, and 8% $Tm^{3+}$-doped nanocrystals. 1064 nm excitation is used, except where noted. See Tables S1,2 for ANP sizes. Photon avalanching is also achieved in the 8% $Tm^{3+}$ ANPs with 1450 nm excitation (brown stars). The dash-dotted lines are fits of the PA DRE model to the data, as described in the text and SI. **b**, 800 nm emission rise times vs. excitation intensity for 8% $Tm^{3+}$ ANPs in (a), showing a large increase, up to 608 ms, near the PA threshold.

PA from other nonlinear multiphoton processes, including conventional energy transfer upconversion (ETU, Fig. 1a(inset)), multistep upconversion via ESA[5,6], and energy looping[30].

Plots of 800 nm $Tm^{3+}$ emission versus 1064 nm pump intensity measured on nanoparticle ensembles drop-casted onto glass substrates (Methods) show that as $Tm^{3+}$ content is increased from 1% to 4%, the degree of nonlinearity ***s*** also increases, but still resides firmly in the energy looping regime, with ***s*** ≤ 7 (Fig. 2a). At these $Tm^{3+}$ concentrations, the chain reaction of ESA and cross-relaxation is too slow to compensate for radiative and multiphonon relaxation from the $^3F_4$ intermediate state, which occurs with rate $W_2$[30]. However, at 8% $Tm^{3+}$ doping, a clear threshold is observed at pump intensity of ca. 20 kW/$cm^2$ (as described in Fig. S3), beyond which the combination of cross-relaxation and ESA act as a gain, and a nonlinear slope ***s*** > 22 is achieved, surpassing the maximum value of 7 observed in the existing pre-avalanching systems (Fig. 2a, green circles). Up- and down-scans of excitation intensity display no measurable photobleaching nor hysteresis, thus showing no significant contribution from excitation-induced thermal avalanching (Fig. S4)[50]. Critically, all three PA criteria are met at room temperature for these 8% $Tm^{3+}$ ANPs (Fig. 2), establishing them as a new class of extremely nonlinear luminescent nanomaterials.

To understand why 8% $Tm^{3+}$ doping gives rise to such non-linear emission, we modeled the PA process in ANPs using coupled nonlinear differential rate equations[26,51] (DREs) (see SI and Tables S3-5 for details). Fitting the model to the experimental data for 8% $Tm^{3+}$ ANPs (Fig. 2a, green dash-dotted line) yields an ESA-to-GSA ($R_2/R_1$) ratio of approximately 13,000 (Table S4), satisfying the $R_2/R_1 > 10^4$ criterion for PA[6,52].

To observe the signature slow-down in excited-state population rise-times expected for PA[4,6,26,53], time-dependent luminescence from the $Tm^{3+}$ $^3H_4$ level (800 nm emission) was measured (Fig. 2b; Methods and

Fig. S5). Rise time is defined as the time needed to reach 95% of the asymptotic value (Fig. S6). We observe that a significant delay of the luminescence rise-time emerges near the PA threshold intensity, reaching a maximum of approximately 608 ms (Fig. 2b) – nearly 400-fold the lifetime of the $^3F_4$ state – further verifying that the PA mechanism prevails in these nanoparticles.

Our modeling of $Tm^{3+}$-based ANPs also predicts PA for even longer-wavelength excitation near 1450 nm, resonant with ESA between $^3F_4$ and $^3H_4$ but not with GSA (Fig. 1c). This is a technologically attractive wavelength range as it is beyond the absorption cutoff of Si-based detectors while leading to emission easily detected by Si, and is also useful for deep-tissue imaging, including through-skull fluorescence imaging of live mouse brain at depths >2 mm[54]. Using pulsed 1450 nm excitation, we indeed observe PA, with the emission versus intensity curve showing a threshold at ~40 kW/cm$^2$ and maximum nonlinearity $s$ = 14.9 (Fig. 2a, brown stars). More generally, the ANPs demonstrate PA for wavelengths between 1400 nm and 1470 nm (Fig. S7), with the lowest threshold occurring at 1450 nm in this range.

Recent theoretical treatments show that achieving PA with a large nonlinearity involves a complex balance between several coexisting phenomena within the material[7]. But in the limiting case where the cross-relaxation rate $s_{31} \gg W_2$, the DRE model predicts that threshold intensity is determined entirely by $W_2$[5,26]. In ANPs, $s_{31}$ is controlled by $Ln^{3+}$ concentration, while the nonradiative decay component of $W_2$ is dominated by losses at surfaces and interfaces[38,44,45,55,56]. To determine if rebalancing these factors would reduce threshold intensity, we synthesized two new 8% $Tm^{3+}$ core/shell structures designed to reduce surface losses and thus $W_2$. These designs include thicker shells as well as larger core size, to further reduce the surface-to-volume ratio, than the 8% ANPs in Fig. 2. The changes indeed result in a distinct reduction in threshold, to <10 kW/cm$^2$ at room temperature (Fig. 3a, top panel). Average core diameter and shell sizes are given in Fig. 3a. We note that the larger cores are slightly prolate in shape (Fig. S2).

We further hypothesized that increasing the $Tm^{3+}$ content should change $s_{31}$ and $W_2$, and therefore the PA excitation threshold intensity. To study this effect, core/shell ANPs with 20% and 100% $Tm^{3+}$ were synthesized, and threshold intensity is found to increase with increasing $Tm^{3+}$ content (Fig. 3a, bottom panel). This is consistent with recent studies showing that, at these pump intensities, excited-state lifetimes are reduced ($W_2$ is increased) as Ln content increases within nanoparticles, with the resulting increase in ion-ion ET opening many potential relaxation pathways that act collectively to depopulate and repopulate the levels[38,57].

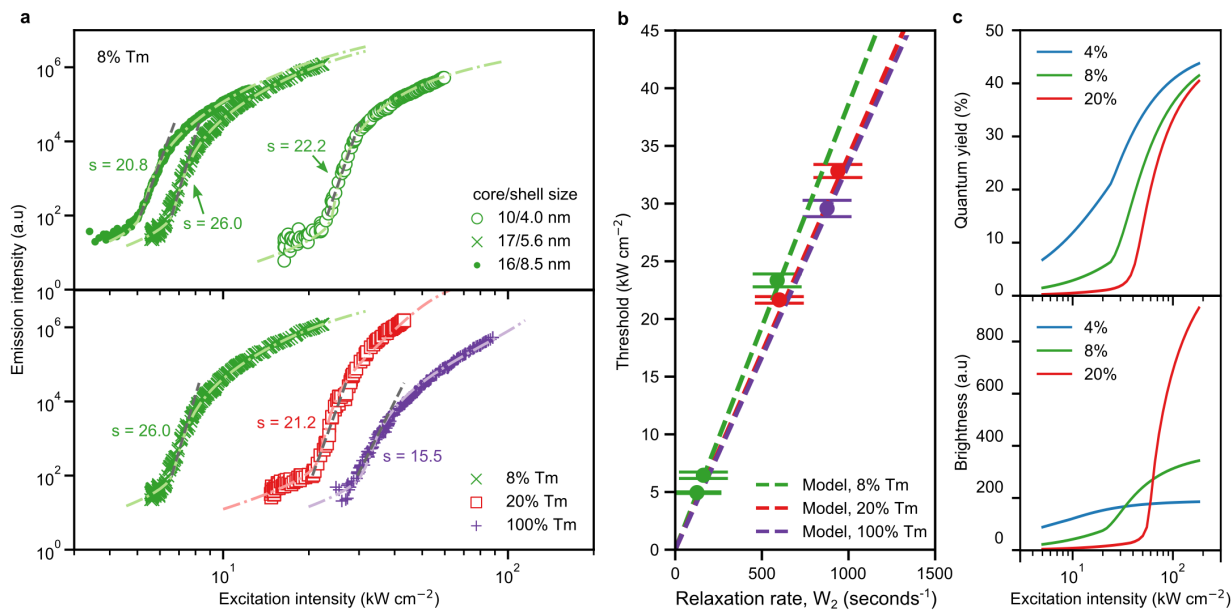

**Fig. 3. Modifying PA kinetics via ANP shell thickness, surface-to-volume ratio, and Tm$^{3+}$ content. a,** top panel. 800 nm emission intensity vs. 1064 nm excitation intensity curves for different core sizes/shell thicknesses of 8% Tm$^{3+}$-doped ANPs. bottom panel: ANPs with different Tm$^{3+}$ concentrations. Green × symbols: 8% Tm$^{3+}$, same as top panel. Red squares: 20% Tm$^{3+}$. Purple + symbols: 100% Tm$^{3+}$. See SI Tables S1,2 for measured dimensions and their standard deviations. The dash-dotted lines are fits of the PA DRE model to the data. **b,** Plot of threshold intensity vs. $W_2$ extracted from the data in (a), showing linear dependencies on $W_2$, with slopes that depend on $s_{31}$. **c,** calculations of upconverting quantum yield and **d,** brightness vs. excitation intensity for 4%, 8%, and 20% Tm$^{3+}$, using values from model fits to the green circles and red squares in (**a**), and the blue circles in Fig. 2a.

Models predict a linear dependence between PA threshold intensity and $W_2$, with a slope that is determined by $s_{31}$, $W_3$ (the excited-state decay rate; see Fig. 1c), and the excited-state relaxation branching ratio[5,26]. These dependencies are shown in Fig. 3b for three different Tm$^{3+}$ concentrations. As $s_{31}$ increases, $W_3$ and the branching ratio become less important, leading to a slight reduction in slope in the threshold intensity-$W_2$ curves. The presence of the 20% and 100% Tm$^{3+}$ data points on nearly the same line demonstrates that by the time Tm$^{3+}$ content reaches 20%, $s_{31}$ dominates and the relative effects of $W_3$ and the branching ratio become almost negligible. This well-defined relationship between the PA threshold and $W_2$ shown in Fig. 3b has important implications for sensing applications, where $W_2$ can be modulated by environmentally dependent ET to the ANP surface, with small changes in $W_2$ (and thus threshold) resulting in large changes in luminescence for a given pump intensity.

To evaluate the efficiency and relative brightness of ANPs, both of which are key considerations for future applications, we used a kinetic computational model of ET within Ln$^{3+}$-doped nanoparticles, similar to those used to reproduce the experimental upconverting quantum yields (QYs) of Er$^{3+}$/Yb$^{3+}$-doped UCNPs[42,58], as well as ELNPs[30]. Our calculations reveal that for fully passivated core-shell nanoparticles,

QY can exceed 20% for ANPs excited beyond threshold at $10^5$ W/cm$^2$ (Fig. 3c). While the model has known limitations – in particular, the absence of higher-energy excited states – we note that calculated QYs are consistent both with previous QY calculations for ELNPs[30] and QY measurements of PA-induced upconversion in fibers at room temperature[21]. In our calculations, we find that while the 8% Tm$^{3+}$ ANPs are somewhat more efficient than 20% ANPs at this pump fluence, the 20% ANPs are brighter (Fig. 3c). This is because brightness is a function of QY, but also the total number of emitters within the particle (brightness is defined as the product of the wavelength-dependent Tm$^{3+}$ ion absorption cross-section, the Tm$^{3+}$ concentration, and QY). Note that the emission intensity shows a more nonlinear dependence on pump fluence than does QY, since the extreme nonlinearity of PA emission is a function of both intensity-dependent QY and excited-state populations.

A particularly compelling application for ANPs is single-particle superresolution imaging, as elucidated by the recently proposed photon-avalanche single-beam superresolution imaging (PASSI) concept that exploits the extreme nonlinear response of PA[7]. Because the size of the imaging point spread function in scanning confocal microscopy (SCM) scales inversely with the square root of the degree of nonlinearity $s$ (as in multiphoton microscopy)[7], deeply sub-wavelength resolution would be realized automatically with ANPs during standard SCM. The imaging requires no complex instrumentation, excitation beam shaping or patterning, image post-processing, or alignment procedures[7].

We performed single-ANP imaging, measuring a PASSI image spot of ≤75 nm average full width at half maximum (FWHM) when excited at 1064 nm near the PA threshold intensity (see Methods for imaging details). More specifically, the image of the 8% Tm$^{3+}$ ANP, from the batch with $s$ = 26 (Fig. 3a), shows a short-axis FWHM of 65 ± 7 nm and a long-axis FWHM of 81 ± 9 nm (Fig. 4b and S8), with its elliptical shape due to a slightly elliptical excitation spot. This spot size agrees well with PASSI simulations (Fig. 4e). The comparison with a diffraction limited excitation spot size of 357 nm FWHM clearly shows the advantage of the extreme nonlinearity of PA (Fig. 4c). In Fig. 4a, the spot size is ~220 nm FWHM when excited closer to the saturation regime, where the degree of nonlinearity $s$ is significantly lower, as predicted[7] (Fig. 4d). The theoretical resolution limit considering $s$ = 26 is given by $\lambda/(2\ NA\ s^{0.5})$ = 70 nm, in excellent agreement with the measured values. PASSI superresolution and its unique power dependence is readily apparent with two ANPs separated by 300 nm just resolvable when excited near saturation, but easily resolvable for intensities in the steep-slope region of the PA emission vs. pump intensity curve (Fig. 4g,h). The resolution is fully determined by the slope of the emission vs. pump intensity (Fig. 4f) curve, allowing us to select the optimal intensity for imaging for a given ANP architecture once that curve is measured[7]. Beyond PASSI, there are also notable advantages for combining the steeply nonlinear ANPs

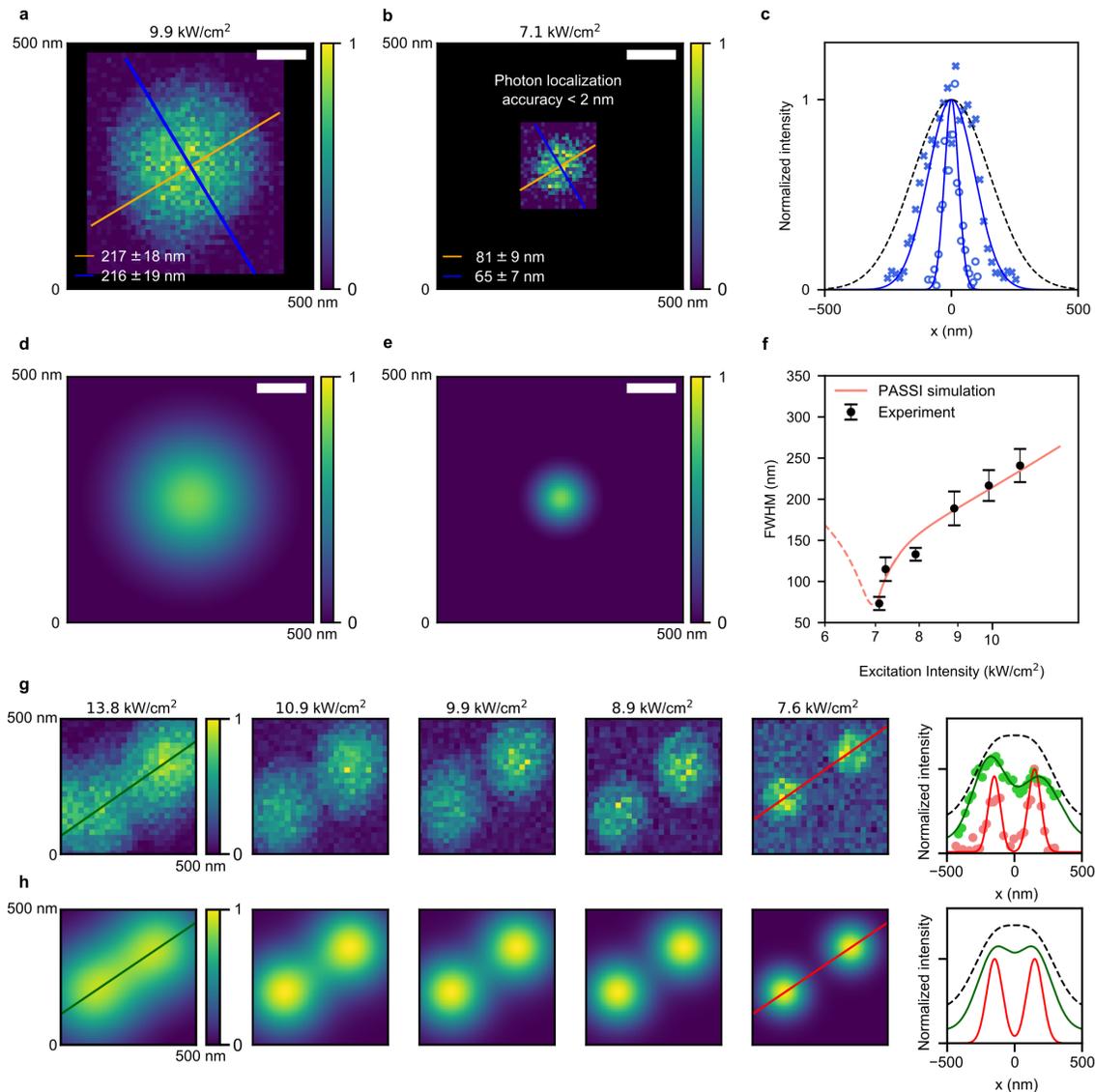

**Fig. 4. Photon-avalanche single-beam superresolution imaging. a-b**, Images of a single 8% $Tm^{3+}$ ANP when excited (**a**) in the saturation regime (9.9 kW/cm$^2$) and (**b**) in the PA regime (7.1 kW/cm$^2$). **c,** linecuts corresponding to the blue lines in (**a**) and (**b**), along with a linecut through a theoretical diffraction-limited focused Gaussian spot (for NA = 1.49, λ = 1064 nm). **d-e**, Simulations of PASSI images for the same excitation intensities in (**a**) and (**b**) based on the measured emission vs. intensity curve shown in Fig. 3a (green x symbols), using the method developed in [7]. **f,** measured (black) vs. simulated (red) FWHMs of single-ANP PASSI images as a function of excitation intensity. The PASSI simulations utilize values from the experimentally measured emission vs. intensity curve shown in Fig. 3a (green x symbols). **g,** Experimental PASSI images of 8% $Tm^{3+}$ ANPs, separated by 300 nm, excited at decreasing intensities, from near saturation (*left*) to near threshold (*right*). Linecuts from the color-coded lines in the images, along with a linecut through a theoretical diffraction-limited image of linear emission from two emitters spaced by 300 nm (black dashed line) (*far right*). **h,** Same as (**g**), but for PASSI simulations.

with existing superresolution approaches. For example, the extreme nonlinearity and anti-Stokes luminescence fundamentally should improve the achievable signal-to-noise and resolution limits of methods such as nonlinear structured illumination microscopy (SIM) and near-infrared emission saturation

(NIRES)[59] nanoscopy for a given photon budget[10,11]. Additionally, applying the photon localization accuracy concept to PASSI images such as that shown in Fig. 4b, which already exhibit sub-100 nm resolution, yields a localization accuracy of <2 nm for only 7600 collected photons, compared to the 10-40 nm accuracies typically achieved[9]. Realizing that the longer rise times might limit scan rates[60], we also calculated a multi-point excitation scheme (Figs. S9,10;), which suggests possible scan rates of approximately 1 frame per second are achievable and reasonable using multi-point PASSI.

Finally, we note that in characterizing this PA system, we measure ~500-10,000-fold increases in emission intensity when pump intensity is increased from threshold to twice the threshold value, which takes us beyond the steep-slope region of the ANP response curve (Figs. 2a,3a). This parameter, which we define as $\Delta_{av} = I_E(2I_P^{th})/I_E(I_P^{th})$, is substantially larger than in reported energy-looping systems (e.g., $\Delta_{av} \leq 50$; refs. 12,30) and suggests a simpler empirical method of identifying PA using a single measurable ratio. $\Delta_{av}$ captures the complex balance between $R_2/R_1$, cross-relaxation, and radiative vs non-radiative relaxation[7]. We find that all nanoparticles with $\geq 8\%$ $Tm^{3+}$ content reported here meet this criteria (Table S6), with a maximum value of ~10,000 attained with 20% $Tm^{3+}$ ANPs, while a borderline value of ~500 is seen in the 100% $Tm^{3+}$ ANPs, where the large increase in cross-relaxation rates leads to faster nonradiative depopulation of $^3H_4$ (ref. 57).

In conclusion, we report a new class of steeply nonlinear nanomaterials, realizing PA in engineered nanocrystals at room temperature with continuous wave pumping. We observe that core-shell architectures doped with only $Tm^{3+}$ ions exhibit avalanching behavior for concentrations $\geq 8\%$, and that the PA excitation threshold intensity is fully determined by the $^3F_4$ intermediate state lifetime at higher concentrations. Further, we show that PA is achieved for excitation in the 1400 – 1470 nm regime in addition to 1064 nm. Along with emission intensities that scale nonlinearly with pump intensity up to the 26th power, (versus maximum nonlinearities of order 7 in current $Ln^{3+}$-based nanosystems[12,32,36]), these results immediately open new applications in ultrasensitive IR photon detection and conversion, local environmental and chemical reporting, and superresolution imaging. More generally, by moving beyond the standard mechanisms, PA can not only enhance upconverting brightness and efficiency, but also offers unprecedented opportunities for technological innovation, since extremely nonlinear materials with controllable properties are essential for new electronic, photonic, and energy management technologies needed to address the growing societal demands for rapid and energy efficient sensing, information processing and transduction.

**References:**

1 Turcotte, D. L. Self-organized criticality. *Rep. Prog. Phys.* **62**, 1377-1429 (1999).
2 Ni, H. & Rand, S. C. Avalanche phase conjugation. *Opt. Lett.* **17**, 1222-1224 (1992).
3 Chivian, J. S., Case, W. E. & Eden, D. D. The photon avalanche: A new phenomenon in Pr3+-based infrared quantum counters. *Appl. Phys. Lett.* **35**, 124-125 (1979).
4 Lenth, W. & Macfarlane, R. M. Excitation mechanisms for upconversion lasers. *J. Lumin.* **45**, 346-350 (1990).
5 Joubert, M.-F. Photon avalanche upconversion in rare earth laser materials. *Optical Materials* **11**, 181-203 (1999).
6 Auzel, F. Upconversion and Anti-Stokes Processes with f and d Ions in Solids. *Chemical Reviews* **104**, 139-174 (2004).
7 Bednarkiewicz, A., Chan, E. M., Kotulska, A., Marciniak, L. & Prorok, K. Photon avalanche in lanthanide doped nanoparticles for biomedical applications: super-resolution imaging. *Nanoscale Horizons* **4**, 881-889 (2019).
8 Deng, H., Yang, S., Xiao, S., Gong, H.-M. & Wang, Q.-Q. Controlled Synthesis and Upconverted Avalanche Luminescence of Cerium(III) and Neodymium(III) Orthovanadate Nanocrystals with High Uniformity of Size and Shape. *Journal of the American Chemical Society* **130**, 2032-2040 (2008).
9 Thompson, M. A., Lew, M. D. & Moerner, W. E. Extending Microscopic Resolution with Single-Molecule Imaging and Active Control. *Annual Review of Biophysics* **41**, 321-342 (2012).
10 Gustafsson, M. G. L. Nonlinear structured-illumination microscopy: Wide-field fluorescence imaging with theoretically unlimited resolution. *Proceedings of the National Academy of Sciences of the United States of America* **102**, 13081 (2005).
11 Heintzmann, R. & Huser, T. Super-Resolution Structured Illumination Microscopy. *Chemical Reviews* **117**, 13890-13908 (2017).
12 Denkova, D., Ploschner, M., Das, M., Parker, L. M., Zheng, X., Lu, Y., Orth, A., Packer, N. H. & Piper, J. A. 3D sub-diffraction imaging in a conventional confocal configuration by exploiting super-linear emitters. *Nat. Commun.* **10**, 3695 (2019).
13 Liu, Y., Wang, F., Lu, H., Fang, G., Wen, S., Chen, C., Shan, X., Xu, X., Zhang, L., Stenzel, M. & Jin, D. Super-Resolution Mapping of Single Nanoparticles inside Tumor Spheroids. *Small* **16**, 1905572 (2020).
14 Marciniak, L., Bednarkiewicz, A. & Elzbieciak, K. NIR–NIR photon avalanche based luminescent thermometry with Nd3+ doped nanoparticles. *Journal of Materials Chemistry C* **6**, 7568-7575 (2018).
15 Kilbane, J. D., Chan, E. M., Monachon, C., Borys, N. J., Levy, E. S., Pickel, A. D., Urban, J. J., Schuck, P. J. & Dames, C. Far-field optical nanothermometry using individual sub-50 nm upconverting nanoparticles. *Nanoscale* **8**, 11611-11616 (2016).
16 Pickel, A. D., Teitelboim, A., Chan, E. M., Borys, N. J., Schuck, P. J. & Dames, C. Apparent self-heating of individual upconverting nanoparticle thermometers. *Nat. Commun.* **9**, 4907 (2018).
17 Lay, A., Sheppard, O. H., Siefe, C., McLellan, C. A., Mehlenbacher, R. D., Fischer, S., Goodman, M. B. & Dionne, J. A. Optically Robust and Biocompatible Mechanosensitive Upconverting Nanoparticles. *ACS Central Science* **5**, 1211-1222 (2019).
18 Zhai, Y., Zhou, Y., Yang, X., Wang, F., Ye, W., Zhu, X., She, D., Lu, W. D. & Han, S.-T. Near infrared neuromorphic computing via upconversion-mediated optogenetics. *Nano Energy* **67**, 104262 (2020).

19  Bradac, C., Johnsson, M. T., Breugel, M. v., Baragiola, B. Q., Martin, R., Juan, M. L., Brennen, G. K. & Volz, T. Room-temperature spontaneous superradiance from single diamond nanocrystals. *Nat. Commun.* **8**, 1205 (2017).

20  Asenjo-Garcia, A., Kimble, H. J. & Chang, D. E. Optical waveguiding by atomic entanglement in multilevel atom arrays. *Proceedings of the National Academy of Sciences* **116**, 25503 (2019).

21  Xie, P. & Gosnell, T. R. Room-temperature upconversion fiber laser tunable in the red, orange, green, and blue spectral regions. *Opt. Lett.* **20**, 1014-1016 (1995).

22  Auzel, F., Chen, Y. & Meichenin, D. Room temperature photon avalanche up-conversion in Er-doped ZBLAN glass. *J. Lumin.* **60-61**, 692-694 (1994).

23  Auzel, F. & Chen, Y. Photon avalanche luminescence of $Er^{3+}$ ions in $LiYF_4$ crystal. *J. Lumin.* **65**, 45-56 (1995).

24  Gomes, A. S. L., Maciel, G. S., de Araújo, R. E., Acioli, L. H. & de Araújo, C. B. Diode pumped avalanche upconversion in $Pr^{3+}$-doped fibers. *Optics Communications* **103**, 361-364 (1993).

25  Martín, I. R., Rodríguez, V. D., Guyot, Y., Guy, S., Boulon, G. & Joubert, M. F. Room temperature photon avalanche upconversion in $Tm^{3+}$-doped fluoroindate glasses. *J. Phys.: Condens. Matter* **12**, 1507-1516 (2000).

26  Guy, S., Joubert, M. F. & Jacquier, B. Photon avalanche and the mean-field approximation. *Phys. Rev. B* **55**, 8240-8248 (1997).

27  Wang, Q.-Q., Han, J.-B., Guo, D.-L., Xiao, S., Han, Y.-B., Gong, H.-M. & Zou, X.-W. Highly Efficient Avalanche Multiphoton Luminescence from Coupled Au Nanowires in the Visible Region. *Nano Lett.* **7**, 723-728 (2007).

28  Ma, Z., Yu, Y., Shen, S., Dai, H., Yao, L., Han, Y., Wang, X., Han, J.-B. & Li, L. Origin of the Avalanche-Like Photoluminescence from Metallic Nanowires. *Scientific Reports* **6**, 18857 (2016).

29  Liu, Y., Lu, Y., Yang, X., Zheng, X., Wen, S., Wang, F., Vidal, X., Zhao, J., Liu, D., Zhou, Z., Ma, C., Zhou, J., Piper, J. A., Xi, P. & Jin, D. Amplified stimulated emission in upconversion nanoparticles for super-resolution nanoscopy. *Nature* **543**, 229-233 (2017).

30  Levy, E. S., Tajon, C. A., Bischof, T. S., Iafrati, J., Fernandez-Bravo, A., Garfield, D. J., Chamanzar, M., Maharbiz, M. M., Sohal, V. S., Schuck, P. J., Cohen, B. E. & Chan, E. M. Energy-Looping Nanoparticles: Harnessing Excited-State Absorption for Deep-Tissue Imaging. *ACS Nano* **10**, 8423-8433 (2016).

31  Fernandez-Bravo, A., Yao, K., Barnard, E. S., Borys, N. J., Levy, E. S., Tian, B., Tajon, C. A., Moretti, L., Altoe, M. V., Aloni, S., Beketayev, K., Scotognella, F., Cohen, B. E., Chan, E. M. & Schuck, P. J. Continuous-wave upconverting nanoparticle microlasers. *Nat. Nanotechnol.* **13**, 572-577 (2018).

32  Si, X., Li, Z., Qu-Quan, W., Hong, D. & Shi-He, Y. Energy Transfer and Avalanche Upconversion of $Nd_x Y_{1-x} VO_4$ Nanocrystals. *Chin. Phys. Lett.* **26**, 124209 (2009).

33  Li, Y., Wang, T., Ren, W., Han, J., Yin, Z., Qiu, J., Yang, Z. & Song, Z. $BiOCl:Er^{3+}$ Nanosheets with Tunable Thickness for Photon Avalanche Phosphors. *ACS Applied Nano Materials* **2**, 7652-7660 (2019).

34  Bednarkiewicz, A. & Strek, W. Laser-induced hot emission in $Nd^{3+}/Yb^{3+}$ : YAG nanocrystallite ceramics. *J. Phys. D: Appl. Phys.* **35**, 2503-2507 (2002).

35  Dwivedi, Y., Bahadur, A. & Rai, S. B. Optical avalanche in $Ho:Yb:Gd_2O_3$ nanocrystals. *J. Appl. Phys.* **110**, 043103 (2011).

36  Wang, G., Peng, Q. & Li, Y. Luminescence Tuning of Upconversion Nanocrystals. *Chemistry – A European Journal* **16**, 4923-4931 (2010).

37  Zhou, B., Shi, B., Jin, D. & Liu, X. Controlling upconversion nanocrystals for emerging applications. *Nat. Nanotechnol.* **10**, 924-936 (2015).

38  Tian, B., Fernandez-Bravo, A., Najafiaghdam, H., Torquato, N. A., Altoe, M. V. P., Teitelboim, A., Tajon, C. A., Tian, Y., Borys, N. J., Barnard, E. S., Anwar, M., Chan, E. M., Schuck, P. J. & Cohen, B.

38  E. Low irradiance multiphoton imaging with alloyed lanthanide nanocrystals. *Nat. Commun.* **9**, 3082 (2018).
39  Tajon, C. A., Yang, H., Tian, B., Tian, Y., Ercius, P., Schuck, P. J., Chan, E. M. & Cohen, B. E. Photostable and efficient upconverting nanocrystal-based chemical sensors. *Optical Materials* **84**, 345-353 (2018).
40  Bünzli, J.-C. G. & Piguet, C. Taking advantage of luminescent lanthanide ions. *Chem. Soc. Rev.* **34**, 1048-1077 (2005).
41  Gnach, A., Lipinski, T., Bednarkiewicz, A., Rybka, J. & Capobianco, J. A. Upconverting nanoparticles: assessing the toxicity. *Chem. Soc. Rev.* **44**, 1561-1584 (2015).
42  Gargas, D. J., Chan, E. M., Ostrowski, A. D., Aloni, S., Altoe, M. V. P., Barnard, E. S., Sanii, B., Urban, J. J., Milliron, D. J., Cohen, B. E. & Schuck, P. J. Engineering bright sub-10-nm upconverting nanocrystals for single-molecule imaging. *Nat. Nanotechnol.* **9**, 300 (2014).
43  Garfield, D. J., Borys, N. J., Hamed, S. M., Torquato, N. A., Tajon, C. A., Tian, B., Shevitski, B., Barnard, E. S., Suh, Y. D., Aloni, S., Neaton, J. B., Chan, E. M., Cohen, B. E. & Schuck, P. J. Enrichment of molecular antenna triplets amplifies upconverting nanoparticle emission. *Nat. Photonics* **12**, 402-407 (2018).
44  Fischer, S., Bronstein, N. D., Swabeck, J. K., Chan, E. M. & Alivisatos, A. P. Precise Tuning of Surface Quenching for Luminescence Enhancement in Core–Shell Lanthanide-Doped Nanocrystals. *Nano Lett.* **16**, 7241-7247 (2016).
45  Johnson, N. J. J., He, S., Diao, S., Chan, E. M., Dai, H. & Almutairi, A. Direct Evidence for Coupled Surface and Concentration Quenching Dynamics in Lanthanide-Doped Nanocrystals. *Journal of the American Chemical Society* **139**, 3275-3282 (2017).
46  Liu, Q., Zhang, Y., Peng, C. S., Yang, T., Joubert, L.-M. & Chu, S. Single upconversion nanoparticle imaging at sub-10 W cm−2 irradiance. *Nat. Photonics* **12**, 548-553 (2018).
47  Chen, X., Jin, L., Kong, W., Sun, T., Zhang, W., Liu, X., Fan, J., Yu, S. F. & Wang, F. Confining energy migration in upconversion nanoparticles towards deep ultraviolet lasing. *Nat. Commun.* **7**, 10304 (2016).
48  Wang, F., Deng, R., Wang, J., Wang, Q., Han, Y., Zhu, H., Chen, X. & Liu, X. Tuning upconversion through energy migration in core–shell nanoparticles. *Nat. Mater.* **10**, 968-973 (2011).
49  Liu, Y., Teitelboim, A., Fernandez-Bravo, A., Yao, K., Altoe, M. V. P., Aloni, S., Zhang, C., Cohen, B. E., Schuck, P. J. & Chan, E. M. Controlled Assembly of Upconverting Nanoparticles for Low-Threshold Microlasers and Their Imaging in Scattering Media. *ACS Nano* **14**, 1508-1519 (2020).
50  Gamelin, D. R., Lüthi, S. R. & Güdel, H. U. The Role of Laser Heating in the Intrinsic Optical Bistability of Yb3+-Doped Bromide Lattices. *The Journal of Physical Chemistry B* **104**, 11045-11057 (2000).
51  Butcher, J. C. *Numerical Methods for Ordinary Differential Equations*.  (Wiley, 2016).
52  Goldner, P. & Pelle, F. Photon avalanche fluorescence and lasers. *Optical Materials* **5**, 239-249 (1996).
53  Joubert, M. F., Guy, S. & Jacquier, B. Model of the photon-avalanche effect. *Phys. Rev. B* **48**, 10031-10037 (1993).
54  Hong, G., Diao, S., Chang, J., Antaris, A. L., Chen, C., Zhang, B., Zhao, S., Atochin, D. N., Huang, P. L., Andreasson, K. I., Kuo, C. J. & Dai, H. Through-skull fluorescence imaging of the brain in a new near-infrared window. *Nat. Photonics* **8**, 723-730 (2014).
55  Ostrowski, A. D., Chan, E. M., Gargas, D. J., Katz, E. M., Han, G., Schuck, P. J., Milliron, D. J. & Cohen, B. E. Controlled Synthesis and Single-Particle Imaging of Bright, Sub-10 nm Lanthanide-Doped Upconverting Nanocrystals. *ACS Nano* **6**, 2686-2692 (2012).

56	Hossan, M. Y., Hor, A., Luu, Q., Smith, S. J., May, P. S. & Berry, M. T. Explaining the Nanoscale Effect in the Upconversion Dynamics of β-NaYF4:Yb3+, Er3+ Core and Core–Shell Nanocrystals. *The Journal of Physical Chemistry C* **121**, 16592-16606 (2017).
57	Teitelboim, A., Tian, B., Garfield, D. J., Fernandez-Bravo, A., Gotlin, A. C., Schuck, P. J., Cohen, B. E. & Chan, E. M. Energy Transfer Networks within Upconverting Nanoparticles Are Complex Systems with Collective, Robust, and History-Dependent Dynamics. *The Journal of Physical Chemistry C* **123**, 2678-2689 (2019).
58	Chan, E. M., Gargas, D. J., Schuck, P. J. & Milliron, D. J. Concentrating and Recycling Energy in Lanthanide Codopants for Efficient and Spectrally Pure Emission: The Case of NaYF4:Er3+/Tm3+ Upconverting Nanocrystals. *The Journal of Physical Chemistry B* **116**, 10561-10570 (2012).
59	Chen, C., Wang, F., Wen, S., Su, Q. P., Wu, M. C. L., Liu, Y., Wang, B., Li, D., Shan, X., Kianinia, M., Aharonovich, I., Toth, M., Jackson, S. P., Xi, P. & Jin, D. Multi-photon near-infrared emission saturation nanoscopy using upconversion nanoparticles. *Nat. Commun.* **9**, 3290 (2018).
60	Pichaandi, J., Boyer, J.-C., Delaney, K. R. & van Veggel, F. C. J. M. Two-Photon Upconversion Laser (Scanning and Wide-Field) Microscopy Using Ln3+-Doped NaYF4 Upconverting Nanocrystals: A Critical Evaluation of their Performance and Potential in Bioimaging. *The Journal of Physical Chemistry C* **115**, 19054-19064 (2011).

**Acknowledgements:** PJS, YDS, SHN and CL gratefully acknowledge support from the Global Research Laboratory (GRL) Program through the National Research Foundation of Korea (NRF) funded by the Ministry of Science and ICT (no. 2016911815), and KRICT (KK2061-23, SKO1930-20). YDS acknowledges the Industrial Strategic Technology Development Program (no. 10077582) funded by the Ministry of Trade, Industry, and Energy (MOTIE), Korea. EX gratefully acknowledges support from the NSF Graduate Research Fellowship Program. Y.L. was supported by a China Scholarship Council fellowship. A.T. was supported by the Weizmann Institute of Science − National Postdoctoral Award Program for Advancing Women in Science. Work at the Molecular Foundry was supported by the Office of Science, Office of Basic Energy Sciences, of the U.S. Department of Energy under Contract No. DE-AC02-05CH11231. KY acknowledges support from Programmable Quantum Materials, an Energy Frontier Research Center funded by the U.S. Department of Energy (DOE), Office of Science, Basic Energy Sciences (BES), under award DE-SC0019443. A.B. acknowledges financial support from NCN, Poland grant number UMO-2018/31/B/ST5/01827.

**Author Contributions**

PJS, EMC, BEC, CL and YDS conceived of the study. Experimental measurements and associated analyses were conducted by CL, EX, YL, AT, KY, AF-B, SHN, and EMC. Advanced nanoparticle synthesis and characterization was performed by YL, AT, and EMC. Theoretical modelling and simulations of photon avalanching photophysics were carried out by CL, EMC, AT, AK and AB. Advanced simulations of superresolution imaging were performed by AK and AB. All authors contributed to the preparation of the manuscript.

**Competing Interests**

The authors declare no competing interests.

**Additional Information**

Supplementary information is available in the online version of the paper. Reprints and permission information is available online at www.nature.com/reprints. Correspondence and requests for materials should be addressed to P. J. S., E. M. C., B. E. C., A. B., and Y.D.S.

**Data Availability**

The data that support the plots within this paper and other findings of this study are available from the corresponding authors upon reasonable request.

**Code Availability**

The code for modelling the PA behavior using the differential rate equations described in the SI are available from the corresponding authors upon reasonable request.